\documentclass[a4paper,12pt]{revtex4}

\usepackage{mathtext}
\usepackage{amsmath}
\usepackage{amstext}
\usepackage{amssymb}
\usepackage{mathrsfs}
\usepackage{graphicx}
\usepackage{wrapfig}
\usepackage{epstopdf}

\def\be#1 {\begin{equation}\label{#1}}
\def\ee{\end{equation}}

\def\d{\partial}

\def\bA{{\bf A}}
\def\bB{{\bf B}}
\def\bE{{\bf E}}
\def\bP{{\bf P}}
\def\bV{{\bf V}}
\def\bx{{\bf x}}
\def\by{{\bf y}}
\def\bS{{\bf S}}

\begin{document}

\title{Commutator of Electric Charge and Magnetic Flux}
\author{Mikhail A. Savrov}
\affiliation{(The Department of Physics and Astronomy, University of New Mexico,\\
on leave from: The Department of General Physics, Moscow Institute of Physics and Technology)}
\date{\today}

\begin{abstract}
Here one can find a derivation of the commutator of the charge and magnetic flux in a superconducting circuit containing a Josephson junction from the commutation relation of quantum-field operators of electric field and vector potential. A commutator of the fluxes of electric and magnetic fields passing through two interlinked loops in vacuum is also evaluated and physical interpretation of the commutator is discussed. 
\end{abstract}

\maketitle

\section{Commutator of Electric Charge and Magnetic Flux in Superconductor}\label{sec1}

Electric circuits with Josephson junctions are macroscopic quantum systems which energy spectrum can contain discrete levels. Such circuits are considered promising for implementation of quantum computing~\cite{bib4}. Theoretical description of the circuits~\cite{bib2} incorporates the commutator of the operators of magnetic flux~$\Phi$ through a circuit loop with a Josephson junction and electric charge~$Q$ on the junction terminals regarded as capacitor plates~\footnote{In Gaussian units; here~$c$ is the speed of light. Often an equivalent commutation relation for~$Q$ and the Josephson phase difference~$\displaystyle{\delta=2\pi\frac{\Phi}{\Phi_0}}$ is used instead of~(\ref{2}). Here~$\Phi_0$ is the magnetic flux quantum.}:
\be2
\left[\Phi,Q\right]=-i\hbar c.
\ee

This commutator has been around for a long time, probably, since the advent of quantum theory. One would inevitably come upon it by trying to quantize oscillations in $LC$-loop! There is an extensive literature in which one can find more or less the same derivation of the commutator~(\ref{2}) as the one  presented here, see e.g. ~\cite{bib4} and~\cite{bib5}, to mention just a few. Nevertheless, it would probably be worthwhile to post a detailed derivation of~(\ref{2}) on the arXive because it is of methodical interest by itself. In addition, the derivation sheds some light on the physical interpretation of the commutator~(\ref{34}) which is usually given only a formal treatment in quantum-field theory textbooks (see section~\ref{sec5}).

Consider the canonical commutation relations of operators of electric field~$\bE$ and vector potential~$\bA$~\cite{bib3},
\be34
\left[E_i(\bx),A_j(\by)\right]=4\pi i \hbar c\, \delta_{ij}\,\delta^3(\bx-\by).
\ee 
This is just the commutator of the canonical coordinate~$\bA$ of electromagnetic field and the canonical momentum~$\bP$ conjugate to~$\bA$, since $\bP=-\bE/4\pi c$~\cite{bib3}. The delta-function on the RHS is due to field nature of the operators. 

Notice that~(\ref{34}) is valid only when~$\bE$ and~$\bA$ are \emph{unconstrained} variables. This is true for superconductor since its minimum energy is attained at~$\bA=(\hbar c/2e) \nabla \theta$, where~$\theta$ is the condensate phase. Therefore, in a superconductor there is no invariance under the gauge transformation
\be40
\bA\rightarrow\bA+\nabla f,
\ee
where~$f$ is an arbitrary function of coordinates, and electromagnetic field (at low energies) has both transverse and longitudinal components~\footnote{Recall that an arbitrary vector field $\bV$ can be decomposed as a sum of longitudinal $\bV_\|$ and transversal $\bV_\perp$ components:
$$
\bV=\bV_\|+\bV_\perp,\qquad\hbox{where}\qquad
\nabla\times\bV_\|=0\quad\hbox{and}\quad\nabla\cdot\bV_\perp=0.
$$}.

\begin{figure}[h]
\begin{center}
\includegraphics[height=3cm]{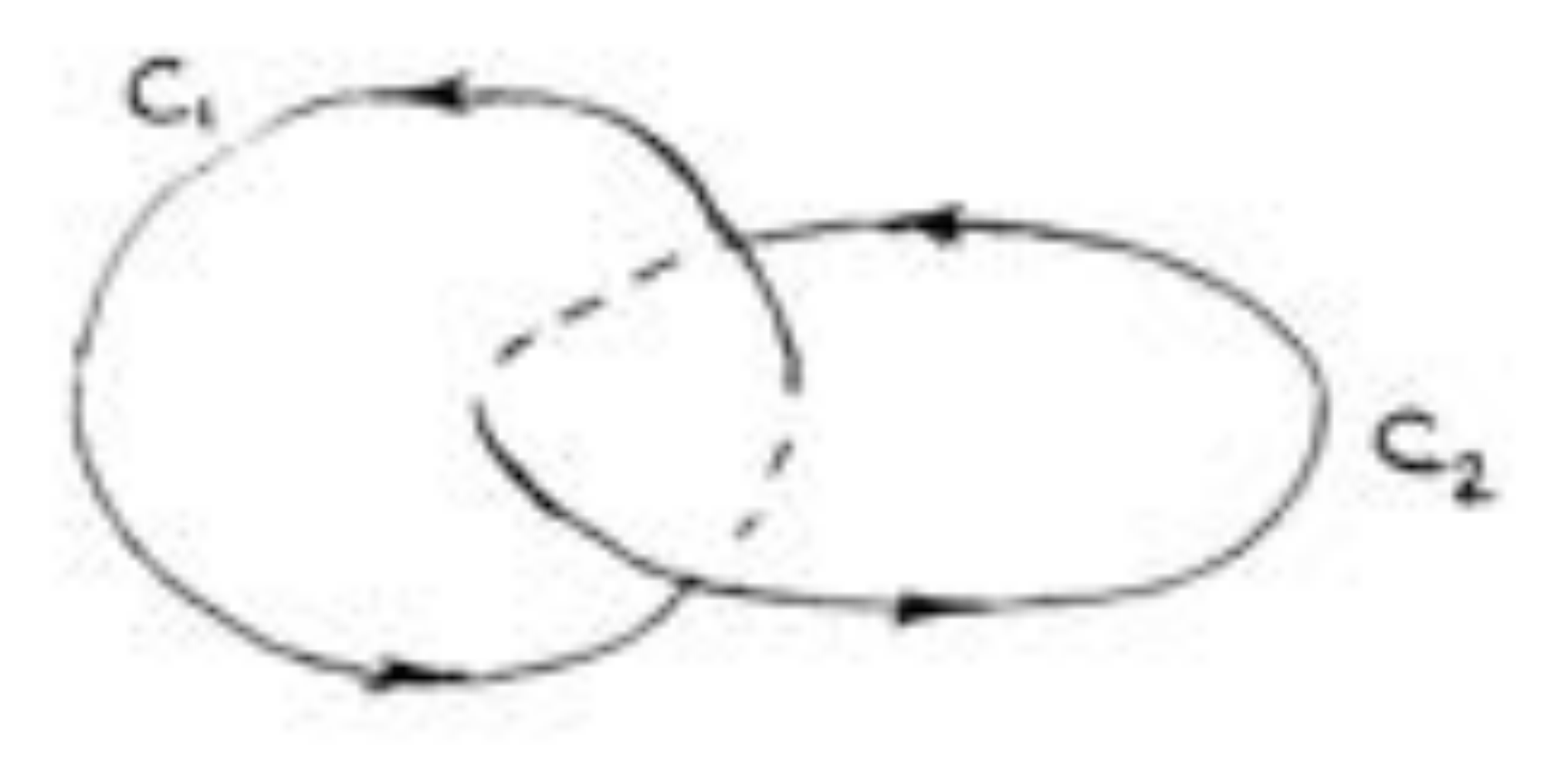}
\caption{ }
\label{fig1}
\end{center}
\end{figure}

Now consider two interlinked loops~$C_1$ and~$C_2$ as shown in Fig.~\ref{fig1}. 
\vspace{0cm}
Let~$C_1$ be the boundary of a surface~$S$ (not shown) and assume that~$C_2$ and~$S$ intersect at a single point. For the chosen loop directions the angle between the normal to~$S$ and the tangent vector to~$C_2$ at the intersection is acute. 

Let us integrate the commutator~(\ref{34}). The variable~$\bx$ is integrated over~$S$, and~$\by$ is integrated around~$C_2$. This gives:
\be4
\left[\int_{S} \bE(\bx)d^2\bS,\oint_{C_2}\bA(\by)d\by\right]=
4\pi i \hbar c\, \int_{S}d^2\bS\oint_{C_2}d\by\,\delta^3(\bx-\by).
\ee
According to Stokes' theorem, the second integral on the LHS of~(\ref{4}) equals a magnetic flux~$\Phi_B$ through an arbitrary surface with the boundary~$C_2$. Therefore, the LHS of~(\ref{4}) is equal to the commutator  $[\Phi_E,\,\Phi_B]$ of electric and magnetic fluxes through surfaces with the boundaries~$C_1$ and~$C_2$, respectively. 

Consider the integral on the RHS of~(\ref{4}). Obviously, the delta-function is nonzero only at the intersection of~$C_2$ and~$S$. To evaluate the integral, let us introduce Cartesian coordinates $(x_1,x_2,x_3)$ so that the intersection point is at the origin and the normal to~$S$ at the intersection coincides with~$\hat{x}_3$. In a small neighbourhood a smooth surface can be considered as approximately flat, therefore coordinates of an arbitrary point of $S$ near the intersection are~$(x_1,x_2,0)$. A small segment of~$C_2$ near the intersection can be represented by a straight line parameterised as $(y\cos\theta_1,y\cos\theta_2,y\cos\theta_3)$ where~$y$ is a distance from the origin and~$\theta_i$ is an angle between the tangent vector to~$C_2$ and~$\hat{x}_i$. In these coordinates the integral on the RHS of~(\ref{4}) near the intersection point becomes: 
\be10
\int dx_1dx_2\int dy\cos\theta_3
\delta(x_1-y\cos\theta_1)\delta(x_2-y\cos\theta_2)\delta(y\cos\theta_3).
\ee
Integration over~$y$ and then over~$x_1,x_2$ reduces the integral to
\be11
\frac{\cos\theta_3}{|\cos\theta_3|}
\int dx_1dx_2\delta(x_1)\delta(x_2)
=\hbox{sgn}(\cos\theta_3).
\ee
By the assumption, the angle between the normal to~$S$ and the tangent vector to~$C_2$ is acute, so~$\cos\theta_3\geq0$. Therefore, the integral on the RHS of~(\ref{4}) equals~+1.

Now let us deform~$S$ with its boundary~$C_1$ (and~$C_2$) held fixed.  In general case, this would produce several intersection points of~$S$ and~$C_2$ and~(\ref{11}) will be replaced by the sum 
\be12
\sum_n\hbox{sgn}(\cos\theta_n)
\ee
over all the intersection points. Here~$\theta_n$ is the angle between the normal to~$S$ and the tangent vector~$C_2$ at the intersection point~$n$. It can be argued that~(\ref{12}) remains equal to~+1. Indeed, if there is a single intersection point, the angle~$\theta$ is acute and the sum is~+1. If there are several intersections, the number of piercings of~$S$ by~$C_2$ from inside out (according to the orientation of~$S$) always exceeds the number of piercings from outside in by one~\footnote{This statement is fairly obvious although a rigorous proof requires topological methods.}. Therefore, the number of acute angles~$\theta_n$ (the sign=+1) exceeds the number of obtuse angles~$\theta_n$ (the sign=-1) by one. Thus,
$$
\int_{S}d^2\bS\oint_{C_2}d\by\,\delta^3(\bx-\by)=1
$$
and
\be{35}
\left[\Phi_E,\Phi_B\right]=4\pi i \hbar c.
\ee

\begin{figure}[h]
\begin{center}
\includegraphics[height=6cm]{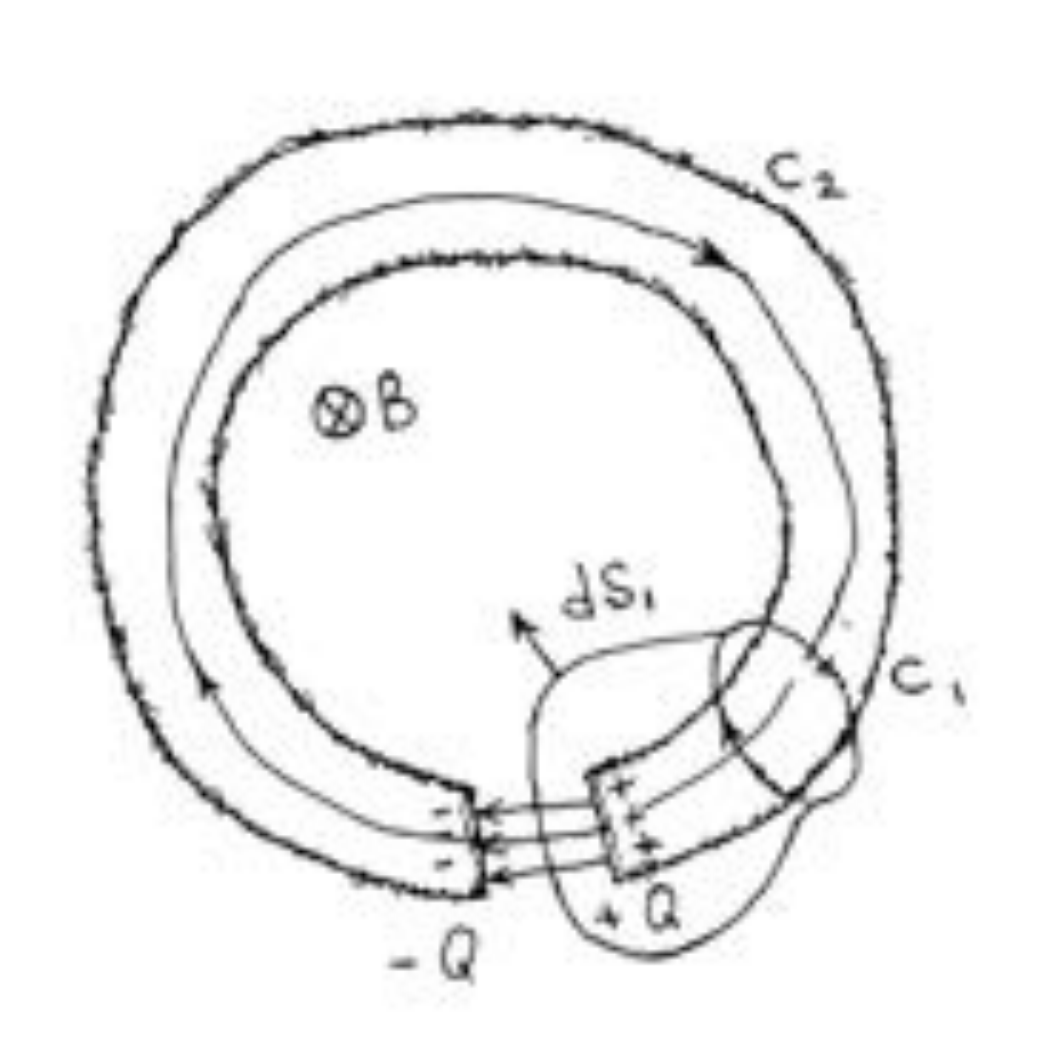}
\caption{Calculation of commutator~(\ref{2}).}
\label{fig2}
\end{center}
\end{figure}

The relation~(\ref{2}) immediately follows from this commutator (see Fig.~\ref{fig2}). Let us draw the loop~$C_2$  
\vspace{-0cm}
along the superconductor axis where there is no magnetic field, so a magnetic flux $\Phi_B$ through the superconductor loop is encircled by~$C_2$. The loop~$C_1$ encloses the superconductor away from the Josephson junction where the electric field is negligible. The surface~$S$ is chosen so that a junction terminal is enveloped by~$S$~\footnote{The intersection point of $S$ and $C_2$ is located within the dielectric, i.e. outside of the superconductor. Nevertheless tunneling of electrons across the junction implies continuity of the wave function of Cooper pairs, hence, continuity of Josephson phase inside the junction and the ensuing longitudinal component of vector potential.}. According to Gauss's theorem, the electric field flux~$\Phi_E$ through~$S$ equals the charge~$Q$ on the junction terminal. Substituting $\Phi_E=4\pi Q$ to~(\ref{35}) and interchanging the operators, one obtains~(\ref{2}).\\

\section{Commutator $\left[\Phi_E,\Phi_B\right]$ in vacuum}\label{sec5}

Operators of electromagnetic field in vacuum have only transverse components and~(\ref{34}) must be accordingly modified~\cite{bib3}:
\be3
\left[E_i(\bx),A_j(\by)\right]_{V}=4\pi i\hbar c\left(\delta_{ij}\delta^3(\bx-\by)
-\frac{\d}{\d x_i}\frac{\d}{\d y_j}\frac{1}{4\pi|\bx-\by|}\right).
\ee
Using the identity
\be19
\Delta\frac{1}{|\bx-\by|}=-4\pi\delta^3(\bx-\by),
\ee
it is easy to verify that applying divergence operator either to $\bE$ or to $\bA$ makes the RHS vanish, so longitudinal components of the fields do not contribute to the commutator.

Let us see how the commutator~$\left[\Phi_E,\Phi_B\right]$ of magnetic and electric fluxes through surfaces with the boundaries~$C_1$ and~$C_2$ (see~Fig.~\ref{fig1}) is modified in vacuum. In addition to the already evaluated integral
\be23
\int_{S}d^2\bS\oint_{C_2}d\by\,\delta^3(\bx-\by)=1,
\ee
there is also a contribution due to the integral
\be18
\oint_{C_2}dy_j\,\frac{\d}{\d y_j}\int_{S}d^2S_i\frac{\d}{\d x_i}\left(-\frac{1}{4\pi |\bx-\by|}\right).
\ee

As in the evaluation of~(\ref{23}), let us first consider the case when~$C_2$ and~$S$ intersect at a single point~$\bx_0$, the loop directions are shown in~Fig.~\ref{fig1}. The surface integral 
\be20
\frac{1}{4\pi}\int_{S}d^{\,2}S_i\left(-\frac{\d}{\d x_i}\frac{1}{|\bx-\by|}\right)
\equiv\frac{1}{4\pi}\Phi(\by)
\ee
is proportional to the electric flux~$\Phi(\by)$ through~$S$ due to a positive unit charge located on~$C_2$ at~$\by$. Next, consider the line integral around~$C_2$:
\be21
\oint_{C_2}dy_j\,\frac{\d}{\d y_j}\Phi(\by).
\ee
The flux is discontinuous at~$\by=\bx_0$, where a flux derivative becomes singular, so~(\ref{21}) is ill-defined. In order to define it, let us discard a small interval which includes~$\bx_0$. The integral~(\ref{21}) defined in this way equals~$4\pi$. Indeed, when the unit charge is just under~$S$ (with respect to the direction of the normal), the direction of the charge electric field on~$S$ coincides with the normal and $\Phi(\bx_0-\vec{0})\rightarrow2\pi$. When the charge is just above~$S$, the field direction on~$S$ is opposite to the normal and $\Phi(\bx_0+\vec{0})\rightarrow-2\pi$. Hence~\footnote{For a more detailed argument leading to~(\ref{22}) see the Appendix.},
\be22
\int^{\bx_0-\vec{0}}_{\bx_0+\vec{0}}dy_j\,\frac{\d}{\d y_j}\Phi(\by)
\rightarrow
\Phi(\bx_0-\vec{0})-\Phi(\bx_0+\vec{0})=4\pi.
\ee
Therefore, the integral~(\ref{18}) equals~+1.

Now let us continuously deform~$S$ with~$C_1$ (and~$C_2$) held fixed. This would change a number of intersections of~$S$ and~$C_2$, however, the number of intersections where the angle between the normal to~$S$ and the tangent vector to~$C_2$ is acute will always exceed by one the number of intersections where the angle is obtuse since such points alternate~\footnote{This is fairly obvious although a rigorous proof requires topological methods.}. The line integral~(\ref{21}) is defined by discarding small segments which include intersection points, so it becomes a sum of integrals over the segments of $C_2$ between the intersections. It is not difficult to verify that the integral over any internal segment vanishes because for such a segment the normals to $S$ at the endpoints are always opposite. Therefore, a continuous deformation of~$S$ does not change the value of~(\ref{18}) which is equal to~+1.  

Thus the sum of~(\ref{23}) and~(\ref{18}) equals~+2 and the commutator of fluxes of electromagnetic field in vacuum is twice of that in superconductor~\footnote{It is a bit counterintuitive that $\left[\Phi_E,\Phi_B\right]$ of purely transversal fields is twice of the commutator of unconstrained fields, but the contribution due to longitudinal components turns out to be negative.}:
\be5
\left[\Phi_E,\Phi_B\right]_{V}=8\pi i\hbar c.
\ee 
If~$C_1$ and~$C_2$ are not interlinked, the commutator vanishes. The fluxes~$\Phi_E$ and~$\Phi_B$ are gauge invariant, so the commutator is independent of the gauge choice. The relation~(\ref{5}) can be regarded as the integral form of field commutator~(\ref{3}) in analogy with integral form of Maxwell's equations.  

The commutator~(\ref{5}) can be interpreted in physical terms. Since the LHS is the imaginary number, this implies the uncertainty relation (the Kennard-Weyl inequality) for the standard deviations of the observables:
\be37
\delta\Phi_E\,\delta\Phi_B\geq4\pi\hbar c.
\ee 
Assume the size of both $C_1$ and $C_2$ be~$\sim l$, then $\delta\Phi_E\sim\delta E\, l^2$ and $\delta\Phi_B\sim\delta B\, l^2$, where~$\delta E$ and~$\delta B$ are fluctuations of $\bE$ and~$\bB$. Then one can write~(\ref{37}) as:
\be38
\frac{1}{4\pi}\delta E\,\delta B\,l^3\geq\frac{\hbar c}{l}.
\ee  
The LHS represents the fluctuation of energy of electromagnetic field in a region of size~$\sim l$ caused by a simultaneous measurement of crossed fields $\bE$ and~$\bB$ (the loops are interlinked) with an accuracy of~$\sim\delta E$ and~$\sim\delta B$. On the RHS there is the energy of a single quantum of a wavelength~$\sim l$. Thus, inequality~(\ref{38}) implies that\emph{ it is impossible to measure  simultaneously and with arbitrary precision the electric and magnetic field of a photon} (crossed fields $\bE$ and~$\bB$ of electromagnetic wave) because the photon itself inevitably vanishes as a result of the measurement. \\

\section{Appendix}\label{sec7}

\begin{figure}[h]
\begin{center}
\includegraphics[height=4cm]{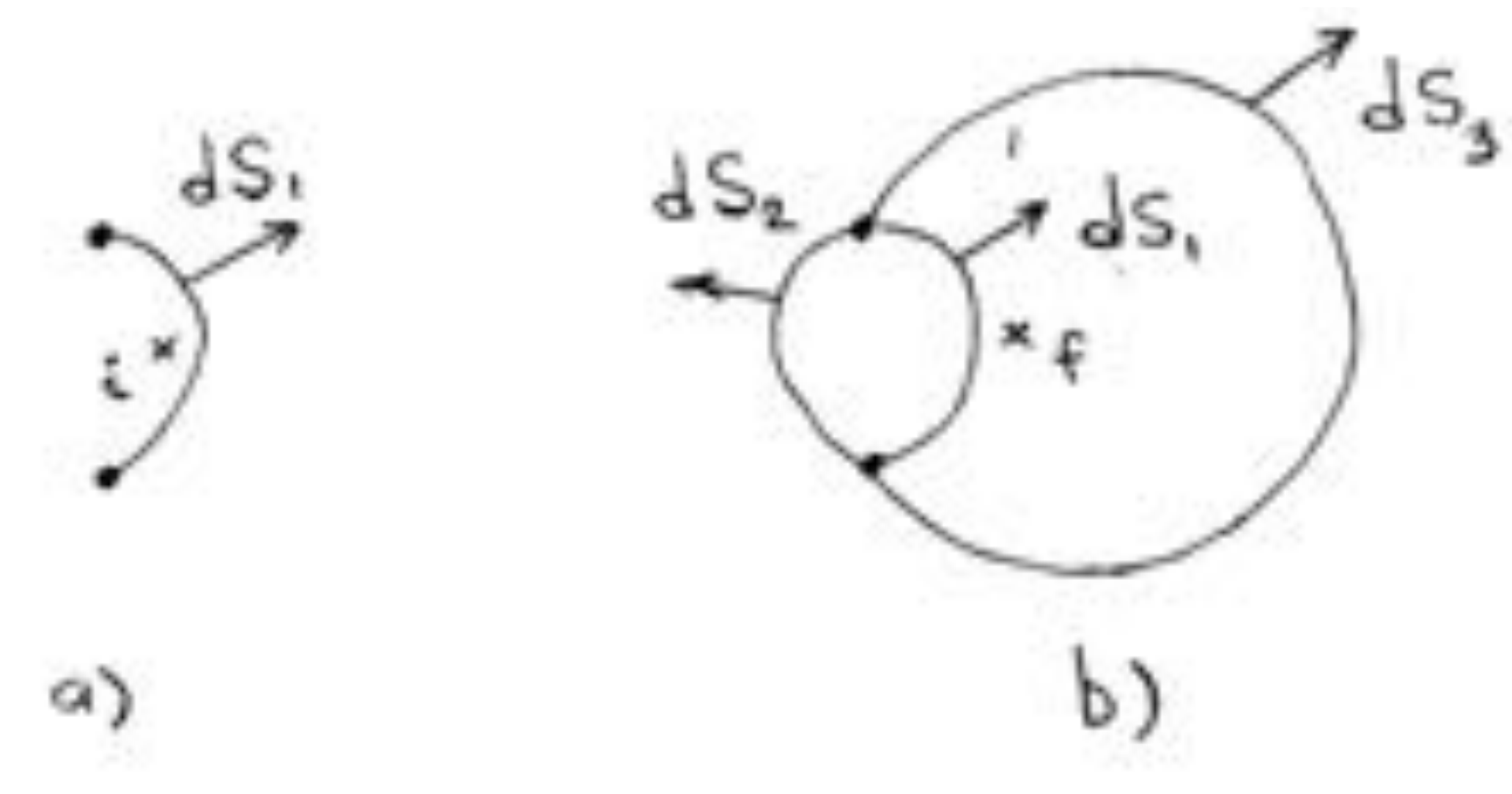}
\caption{Evaluation of~(\ref{22}).}
\label{fig3}
\end{center}
\end{figure}

A more detailed proof of~(\ref{22}) is given below. To this end, let us show that the electric flux through a surface due to a point charge~$Q$ changes by~$-4\pi Q$ when the charge crosses the surface in the direction of the normal to the surface. The diagram in Fig.~\ref{fig3}\,a) shows a cross-section of a surface~$S_1$ with its boundary depicted by two bold dots. A normal to the surface is also shown. The cross~$i$ indicates the charge position before the crossing. 
\vspace{0cm}
Let~$\Phi_i$ be the flux through~$S_1$ when~$Q$ is located at $i$. Let us find the flux~$\Phi_f$ through~$S_1$ when~$Q$ is located at the point $f$ after the crossing. The diagram in Fig.~\ref{fig3}\,b) shows the charge~$Q$ at~$f$, the surface~$S_1$ and its boundary, and two additional surfaces~$S_2$ and~$S_3$ each with the same boundary as $S_1$ and which form a closed surface together. The normals to the surfaces are also shown. Let~$\Phi_2$ and~$\Phi_3$ be the fluxes through~$S_2$ and $S_3$ due to the charge at~$f$. According to Gauss's theorem,
\be13
\Phi_2+\Phi_f=0,\qquad \Phi_2+\Phi_3=4\pi Q.
\ee
Therefore,
\be14
\Phi_3=4\pi Q-\Phi_2=4\pi Q+\Phi_f.
\ee
Since the initial and final positions of the charge are infinitely close, $\Phi_3\rightarrow\Phi_i$. Together with~(\ref{14}) this gives 
\be15
\Phi_f-\Phi_i\rightarrow-4\pi Q.
\ee
To evaluate~(\ref{22}), notice that initial position of the charge~$Q=1$ is above the surface and the final position is below the surface, therefore the sign of~$\Phi_f-\Phi_i$ is reversed.

\section{Acknowledgements}

I am grateful to prof. Morozov~A.~I. who carefully read the manuscript and made valuable remarks, to prof. U.~Vool for endorsing the paper, and especially to prof. Spiller~T.~P. for kindly bringing my attention to the pioneering works in the field of superconducting circuits, where the flux-charge commutator was originally introduced.

\end{document}